\documentclass[12pt, preprint]{aastex}

\def\kms{km s$^{-1}$}
\pdfoutput=1

\def\and{$\&$ }

\usepackage{wrapfig}
\usepackage{graphicx}

\begin{document}

\title{Evidence of Short Timescale Flux Density Variations of UC H~{\sc ii}  regions in Sgr B2 Main and North}

\author{C. G. De Pree\altaffilmark{1}, T. Peters\altaffilmark{2, 3}, M.-M. Mac Low\altaffilmark{4, 8}, D. J. Wilner\altaffilmark{5}, W. M. Goss\altaffilmark{6}, R. Galv\'an-Madrid\altaffilmark{7}, E. R. Keto\altaffilmark{5}, R. S. Klessen\altaffilmark{8} \& A. Monsrud\altaffilmark{1}}

\altaffiltext{1}{Department of Physics \& Astronomy, Agnes Scott College, 141 E. College Ave., Decatur, GA 30030}
\altaffiltext{2}{Institut f\"{u}r Computergest\"{u}tzte Wissenschaften, Universit\"{a}t Z\"{u}rich, Winterthurerstrasse 190, CH-8057 Z\"{u}rich, Switzerland}
\altaffiltext{3}{Max-Planck-Institut f\"{u}r Astrophysik, Karl-Schwarzschild-Str. 1, D-85748 Garching, Germany}
\altaffiltext{4}{Department of Astrophysics, American Museum of Natural History, 79th St. and Central Park West, New York, NY, 10024}
\altaffiltext{5}{Harvard-Smithsonian Center for Astrophysics, 60 Garden St, Cambridge, MA 02138}
\altaffiltext{6}{National Radio Astronomy Observatory, 1003 Lopezville Rd, Socorro, NM 87801}
\altaffiltext{7}{Centro de Radioastronom\'ia y Astrof\'isica, Universidad Nacional Aut\'onoma de M\'exico, Morelia 58090, Mexico}
\altaffiltext{8}{Universit\"{a}t Heidelberg, Zentrum f\"{u}r Astronomie, Institut f\"{u}r Theoretische Astrophysik, Albert-Ueberle-Str. 2, 69120 Heidelberg, Germany}

\begin{abstract}
We have recently published observations of significant flux density variations at 1.3 cm in H~{\sc ii}  regions in the star forming regions Sgr B2 Main and North (De Pree et al. 2014). To further study these variations, we have made new 7 mm continuum and recombination line observations of Sgr B2 at the highest possible angular resolution of the Karl G. Jansky Very Large Array (VLA). We have observed Sgr B2 Main and North at 42.9 GHz and at 45.4 GHz in the BnA configuration (Main) and the A configuration (North). We compare these new data to archival VLA 7 mm continuum data of Sgr B2 Main observed in 2003 and Sgr B2 North observed in 2001. We find that one of the 41 known ultracompact and hypercompact H~{\sc ii}  regions in Sgr B2 (K2-North) has decreased $\sim$27\% in flux density from 142$\pm$14 mJy to 103$\pm$10 mJy (2.3$\sigma$) between 2001 and 2012. A second source, F3c-Main has increased $\sim$30\% in flux density from 82$\pm$8 mJy to 107 $\pm$11 mJy (1.8$\sigma$) between 2003 and 2012. F3c-Main was previously observed to increase in flux density at 1.3 cm over a longer time period between 1989 and 2012 (De Pree et al. 2014). An observation of decreasing flux density, such as that observed in K2-North, is particularly significant since such a change is not predicted by the classical hypothesis of steady expansion of H~{\sc ii} regions during massive star accretion. Our new observations at 7 mm, along with others in the literature, suggest that the formation of massive stars occurs through time-variable and violent accretion.
\end{abstract}

\section{INTRODUCTION}
Accretion flows onto massive stars are so dense that they may become internally gravitationally unstable, leading to the formation of strong density fluctuations in the flows. The interaction of these fluctuations with the ionizing radiation from the massive stars at the centers of the flows may cause stochastic variations in the total mass of ionized gas (the H~{\sc ii}  regions) around the stars (Keto 2002, 2003; Keto \& Wood 2006). In high resolution, radiation-hydrodynamic  and magnetohydrodynamic simulations (Peters et al.\ 2010a, 2010b, 2011), the developing H~{\sc ii} region surrounding a young high mass star flickers between hypercompact (HC, d$< $0.05 pc, EM$>$10$^9$ pc cm$^{-6}$) and ultracompact (UC, d$<$0.1 pc, EM$>$10$^7$ pc cm$^{-6}$) states throughout the main accretion phase.\footnote{There are a number of definitions of UC and HC H~{\sc ii} regions; here we use the criteria of Kurtz (2002).}  

Theories that describe the evolution of H~{\sc ii}  regions around non-accreting stars embedded in a homogeneous medium would predict monotonic expansion. The small size of HC H~{\sc ii}  regions, together with radio recombination line widths indicating velocities of a few tens of kilometers per second (De Pree et al., 2011), suggest time scales on the order of 100 yr, in agreement with the predictions of the simulations (Galv\'an-Madrid et al. 2011).The timescale on which we can expect notable changes in the ionization state in the absence of an ionizing flux is the recombination timescale, which is defined as the inverse of the recombination rate. For 10$^4$ K hot gas, t$_{rec}$ = (2.59 x 10$^{-13}$ n$_e)^{-1}$
  s, when the electron number density n$_{e}$ is given in cm$^{-3}$. For an ultracompact H II region with 
  n$_{e}$ = 10$^4$ cm$^{-3}$, this gives t$_{rec}$ = 12 yr, and for a hypercompact region with n$_e$=10$^6$ cm
  $^{-3}$ we have t$_{rec}$ = 0.12 yr. In reality, of course, the ionizing flux is not shielded perfectly in all directions, the density and temperature in the H II region are inhomogeneous, and collisional ionization also contributes, leading to an order of magnitude estimate of 10 to 100 yr.
 Thus, flux density variations might be detected on observable time scales (10 yr) as changes in the brightness or sizes of the H~{\sc ii}  regions.

Sgr B2 is one of the most luminous star forming regions in the Milky Way, and is associated with a massive giant molecular cloud ($\sim$10$^6$ M$_{\odot}$). The star forming region contains 49 H~{\sc ii} regions (Gaume et al. 1995), of which 35 are ultracompact and 6 are hypercompact.  As the resolution of radio frequency observations has improved, Sgr B2 has been divided up into ever smaller sub regions. Sgr B2 is broadly separated into two areas: Sgr B2 Main, and Sgr B2 North ($\sim$1\arcmin~north). As sub sources were detected, they were given letters in order of increasing RA, and then numbers (e.g. F1a). In this paper, we use the existing naming schemes of Gaume et al. (1995) and De Pree et al. (1998), and in addition append whether the source is located in Main or North (e.g. K2-North). The large number of UC and HC sources within a small field of view makes Sgr B2 ideally suited for testing theories of UC H~{\sc ii} region evolution. Consequently, it has been a frequent target of interferometric observations at radio wavelengths (Gaume et al. 1995, De Pree et al. 1998, Qin et al. 2011, De Pree et al. 2011, De Pree et al. 2014). Gaume et al. (1995) published high resolution ($\theta_{beam}$=0\farcs25, $\sim$2000 AU) 1.3 cm radio images of the Sgr B2 Main and North star forming regions taken in 1989 with the VLA.\footnote{The National Radio Astronomy Observatory is a facility of the National Science Foundation operated under cooperative agreement by Associated Universities, Inc.} De Pree et al. (1998) made the first high-resolution VLA 7 mm continuum observations of Sgr B2 in 1996/7 ($\theta_{beam}$ = 0\farcs065, $\sim$600 AU), and high resolution 7 mm line and continuum observations followed in 2003 (De Pree et al. 2011). 

Fluctuations in radio brightness have been detected in several ionized sources with multi-epoch VLA observations (e.g. Cep A, Hughes 1988; NGC 7538 IRS1, Franco-Hernandez \& Rodriguez 2004; MWC 349A, Rodriguez et al. 2007; G24.78+0.08, Galv\'an-Madrid et al. 2008, Orion BN/KL, Gomez et al. 2008, Dzib et al. 2013, Rivilla et al., 2015). Our previous paper in this investigation (De Pree et al. 2014) presented the changes detected in Sgr B2 Main and North from observations of the 1.3 cm continuum emission between 1989 and 2012. At 1.3 cm, we detected four sources that changed in peak intensity over 23 years. K3-North was detected to decrease in peak intensity by 7\%, and sources F1, F3 and F1a\footnote{Also referred to in the literature as F10.303} were detected to increase in peak intensity by 7\%, 5\% and 16\% respectively. 

In this paper, we report new 7 mm continuum observations of Sgr B2 Main and North. We give 7 mm continuum parameters for all detected sources, including a newly identified 7 mm source in Sgr B2 North (K7-North). Archival observations of Sgr B2 Main and North at 7 mm have allowed us to search for changes in flux density between 2001 and 2012.  In comparing our new observations with archival data, we detect significant changes in flux density from two sources: F3c-Main and K2-North. Simultaneous VLA 7 mm Radio Recombination Line (RRL) observations will be introduced along with new VLA 1.3 cm RRL data from the same regions in a subsequent paper.

\section{VLA OBSERVATIONS and DATA REDUCTION}

\subsection{2012 Sgr B2 Main and North 7 mm line and continuum}
Sgr B2 Main was observed with the VLA at 7 mm on 21 September 2012 (during the move from the BnA to the A configuration), and Sgr B2 North was observed at 7 mm on 28 October 2012 (A configuration) under program name 11B-058. Each observation was for a total of four hours, with $\sim$2.5 hours on source, and the balance of time for flux density, phase and bandpass calibrators. Observations were made at 45.4 GHz and 42.9 GHz, separated into 16 Intermediate Frequencies (IFs). One of the central IFs at each frequency was centered on a radio recombination line (H52$\alpha$ at 45.4 GHz and H53$\alpha$ at 42.9 GHz) with $\sim$450 \kms~bandwidth in each IF. The key observational parameters are summarized in Table 1. 

For the Sgr B2 Main data, the basic calibration was carried out with the new VLA data pipeline, which flags frequencies and antennas with interference, and determines and applies bandpass, phase and amplitude calibrations. The calibrated continuum data were then self-calibrated and imaged using the Common Astronomy Software Applications (CASA). Basic calibration was also carried out for the Sgr B2 North observations using the VLA data pipeline, and self-calibration and imaging were done with the Astronomical Image Processing System (AIPS). The two continuum $uv$-data sets at 7 mm (42.9 and 45.4 GHz) were averaged and then imaged to make a single 44.2 GHz continuum image for Sgr B2 Main and North. Hereafter these 44.2 GHz images images are referred to as Main-2012 and North-2012. 

\subsection{Archival Data (Sgr B2 Main)}
Our previous 7 mm continuum observations of Sgr B2 Main from 1996/7 (43.3 GHz) and 2003 (45.4 GHz) were reported in De Pree et al. (1998) and De Pree et al. (2011) respectively. In order to compare images between epochs with matched uv-coverage, bandwidth and restoring beams, we reprocessed the 1996/7 and the 2003 observations with AIPS. In order to compare source flux densities, it was important to make sure that the uv-coverage and bandwidth were as well-matched as possible, and that the comparison images were made with matched restoring beams. 

In order to have matched uv-coverage between the 1996/7 and the 2012 data, we needed to severely restrict the number of antennas to only the 7 pads that overlapped between the two observations. The 2012 observations were made in the BnA configuration, and the 1996/7 observations were made with the antennas located in a `spiral' pattern\footnote{This pattern was chosen at the time to maximize resolution with a total of only 13 Q-band antennas}. However, the resulting images were of too low quality to make detailed comparisons in the complex region around the F sources.

We had more success making comparison images for the 2012 and the 2003 datasets. Sgr B2 Main was observed with the VLA in the BnA configuration on 4 October 2003. The uv-coverage was well matched between these two observations, as the VLA was in its BnA configuration at both times. To compare the 2003 and the 2012 data, we made an image of the Sgr B2 Main region from a single line-free (32 MHz bandwidth) IF from the 2012 data, and used the same restoring beam for the two images (0\farcs12$\times$0\farcs09, BPA\footnote{Beam Position Angle}=60$^o$). In the discussion that follows, the matched resolution images of Sgr B2 Main are referred to as Main-2003 and Main-2012m.

\subsection{Archival Data (Sgr B2 North)}

Sgr B2 North was observed with the VLA in the BnA configuration on 12 February 2001 (45.7 GHz). These data were edited and calibrated using standard techniques in AIPS, selecting only line-free channels to make the continuum image. In order to compare the 2001 and the 2012 images of Sgr B2 North, we had to restrict the uv-coverage of the 2012 data (which was observed in the A configuration). We set a uv-range when imaging the 2012 data that matched the uv-range in the 2001 data (with a maximum telescope separation of $\sim$2.5 M$\lambda$). We also imaged only a single IF from the 16 IFs of the 2012 data to match the bandwidth in the 2001 data. The resulting images were made with a matched restoring beam of (0\farcs13$\times$0\farcs10, BPA=0$^o$). In the discussion that follows, the matched resolution images of Sgr B2 North are referred to as North-2001 and North-2012m.

\section{RESULTS}

 \subsection{7 mm Continuum in Sgr B2 Main and North (2012)}
The central region of the full bandwidth image (Main-2012) is presented in Figure 1 ($\theta_{beam}$ = 0\farcs12$\times$0\farcs10, BPA=73$^o$, contours). The rms noise in the full resolution image is 0.14 mJy beam$^{-1}$. The dashed box in Fig. 1 indicates the region of detail shown in Fig. 4. Sources are labeled using the naming conventions of De Pree et al. (1998), with sub sources labelled in Fig. 4. 

Figure 2 shows the full bandwidth image (North-2012; $\theta_{beam}$=0\farcs11$\times$0\farcs05, BPA=3$^o$). The rms noise of the image is 0.16 mJy beam$^{-1}$. Sources are labeled using the naming conventions of Gaume et al. (1995). The sources in Sgr B2 North have unique morphologies, apparent for the first time in this high angular resolution image. K1-North has a cometary morphology, with a sharp cutoff in surface brightness on its western edge. K2-North consists of a bright, compact source, with a region of more diffuse, extended emission  to the NW. Source K3-North is shell-like ($D\sim$ 0\farcs25, 0.01 pc), with breaks in the edge-brightened shell to the N and SE. K7-North is double peaked, with the brighter of the two peaks located to the SE. Finally, K4-North is a larger, shell-like source, with a diameter of $\sim$1\arcsec ($\sim$0.04 pc). The dashed box indicates the region of detail in Fig. 3.

For each source above the 3$\sigma$ noise level in the continuum images for Sgr B2 Main and North, we report the following parameters from 2-D Gaussian fits (numbers in square brackets indicate column number): position [1,2], fitted source size [3], peak intensity (mJy beam$^{-1}$) [4], and flux density (Jy) [5]. Table 2a shows these parameters for Main-2012 and Table 2b shows them for North-2012.

\subsection{7 mm Difference Images}
In order to search for flux density changes in Sgr B2 Main and Sgr B2 North, we compared source flux densities in the matched-resolution images. The analysis in Sgr B2 North was considerably simpler, so we begin with this region.

\subsubsection{Sgr B2 North}
Sgr B2 North contains a smaller number of well-isolated sources, so the comparison between epochs was more straightforward. In order to look for flux density changes in North, we first normalized the two matched-resolution North images (North-2001 and North-2012m), aligned the two images, and then subtracted the two matched datasets. To determine the normalization factor, we took the ratio of the fitted flux densities of sources K1-North, K2-North, K3-North and K7-North. The ratio was $\sim$1.3 for sources K1, K3 and K7, and 0.85 for source K2. To normalize the flux density between the images, we multiplied 2001-North by a factor of 1.3 (in AIPS) so that the flux density ratios for sources K1, K3, and K7 would be 1 between the two epochs. 

Fig. 3 shows the results of the difference image in Sgr B2 North. Fig. 3a shows North-2012m in a greyscale, and the difference between the two images indicated as contours. The dashed box indicates the region of detail shown in Fig. 3b and Fig 3c. Fig. 3b shows the continuum in detail around source K2-North and the difference image between the normalized, matched resolution images North-2012m and North-2001 (contours). Finally, Fig 3c shows the 7 mm continuum from K2-North as contours, showing the extension of emission to the NW of the region of flux density change. Table 3 presents the fits to the North-2001 and North-2012m datasets. Continuum source parameters were determined for the sources in Sgr B2 North using the JMFIT task in AIPS. For each source above the 3$\sigma$ noise level in the continuum images for Sgr B2 North, we report the following parameters from 2-D Gaussian fits (numbers in square brackets indicate column number): position [1,2], fitted source size [3], peak intensity (mJy beam$^{-1}$) [4], and flux density (Jy) [5]. Table 3 shows these parameters for North-2012m and North-2001. The reported size of K4-North is an average of two perpendicular slices made across the source in AIPS. K4-North was too faint to be fitted in the 2001 data. 

In Table 3, there are two additional columns that list the ratio of the fitted peak intensities between the most recent observations (North-2012m), and the oldest archival data available for those sources (North-2001) [6], and the ratio of flux densities for the same two epochs [7]. Values from Tables 3 were used to look for changes in peak intensity and flux density discussed below. K2-North is observed to decrease in both peak intensity and flux density. The other 3 detected continuum sources (K1, K3, and K7) are relatively constant. Source K7, newly discovered in the 2012 data, is also detected in the 2001 data presented here.

In order to check that the normalization described above has not spuriously introduced the change detected in K2-North, we note that the flux density ratios of pairs of sources within a single epoch give the same result as the comparison of the normalized images. That is, the flux density ratio of the K7/K3 sources does not change between 2001 and 2012, with K7/K3$_{2001}$ = 0.14$\pm$0.1 and K7/K3$_{2012}$=0.13$\pm$0.1. The ratio of the K7/K1 sources is also constant (within the errors of the ratio), with K7/K1$_{2001}$=0.20$\pm$0.03 and K7/K1$_{2012}$=0.22$\pm$0.03. By contrast, the ratio between K2-North and its companion sources in Sgr B2 North within the same epoch does change between 2001 and 2012. K2/K7$_{2001}$=4.7$\pm$0.7, but this ratio falls to 3.3$\pm$0.4 by 2012. Likewise, the ratio K2/K3$_{2001}$ falls from 0.66$\pm$0.10 to 0.43$\pm$0.06 in 2012. As a result, we conclude that making difference images of the normalized, matched uv-coverage, matched bandwidth images is a valid method to look for individual source flux density changes. Table 3 indicates that source K2-North is the only source in Sgr B2 North to show a decrease in both peak intensity and flux density between 2001 and 2012.

\subsubsection{Sgr B2 Main}
The crowded field in the boxed region shown in Fig. 1 of Sgr B2 Main made Gaussian source fitting more difficult than it was in Sgr B2 North. In Sgr B2 Main, we chose a slightly different method to look for flux density changes. We normalized the 2003-Main image in the same manner, choosing the ratio of the G and F10.37 sources (well-isolated, high signal to noise sources) to determine the normalization factor. For these two sources, the ratio of the 2012 to the 2003 flux density was 0.72$\pm$0.02, so we multiplied the Main-2003 image by 0.72, and then aligned and subtracted the two images.

Fig. 4 shows the region of detail from Fig. 1 of the sources in Sgr B2 Main. Fig. 4a shows the Main-2012 data as contours, with sub sources named as in De Pree et al. (1998). 
Fig 4b shows the Main-2012m image (grayscale) and the difference image between Main-2012m and Main-2003 as contours. The difference image highlights two regions where flux density appears to have increased between 2003 and 2012: F3c-Main and F1f-Main. Only the increase in peak intensity in F3c-Main is greater than 2$\sigma$ (see below).

%Taking the Main-2012 flux densities of F3c-Main and F1f-Main from Table 2, and the change in flux density seen in the difference image, we see that F3c-Main increases in flux density by 24.6 mJy to 107 mJy (a 2.3$\sigma$ increase), and F1f increases by 12.5 mJy to 153 mJy (an $\sim$1$\sigma$ increase).

%Assuming 10\% errors in our flux densities (consistent with flux density errors typical of the VLA, and with the errors of the bootstrapped flux density values for the phase calibrators calculated during data reduction), we see that F3c-Main and F1f-Main appear to have increased in flux density by $\sim$30\% and $\sim$9\% respectively. The ratios of the flux densities can be written as 1.30$\pm$0.19 and 1.09$\pm$0.16 for F3c-Main and F1f-Main respectively. Of these two sources, we can say with confidence that F3c-Main has increased in flux density, while the increase in F1f-Main falls within our error.
 
\subsection{Sources with Flux Density Changes}

In the current data, we have observed significant flux density changes in two sources: K2-North and F3c-Main. These two sources represent $\sim$5\% of the known UC and HC H~{\sc ii} regions in Sgr B2. F3c-Main has increased $\sim$30\% in flux density, from 82$\pm$8 mJy to 107$\pm$11 mJy (1.8$\sigma$). In the Sgr B2 North region,  K2-North has decreased $\sim$29\% in peak intensity from 40.7$\pm$2.0 mJy beam$^{-1}$ to 28.9$\pm$0.3 mJy beam$^{-1}$ (5.9$\sigma$) and $\sim$27\% in flux density from 142$\pm$14 to 103$\pm$10 mJy (2.3$\sigma$). We note that a flux density increase in F3c-Main has also been detected at 1.3 cm (De Pree et al. 2014). In K2-North, the decrease in flux density is slightly offset from the continuum peak. Based on the simulations of Peters et al. (2010a, 2010b, 2011), the detected variations are not expected to necessarily be centrally peaked. The location of the variations depends on the interaction of the radiation with the complex accretion flow (Keto 2002, 2003; Keto \& Wood 2006).

As an additional way to view the changes over time in Sgr B2 North, the source flux densities of the individual sources in North-2001 and North-2012m are shown in Fig. 5. Note that only source K2-North appears to have a significant change in flux density between the two epochs. The other sources are consistent with level flux densities between the two epochs.

\subsubsection{Observations at Different 7 mm Frequencies}
The observations at 7 mm were all made at slightly different frequencies, ranging from 44.2 GHz (Main-2012) to 45.4 GHz (Main-2003) to 45.7 GHz (North-2001). Using the relationship $S_\nu\propto\nu^{\alpha}$ (where S$_{\nu}$ is continuum flux density and $\alpha$ is spectral index), we can examine whether any of the flux density changes detected at 7 mm could result from comparing observations at different 7 mm (Q-Band) wavelengths. The range in spectral index values possible in UC and HC H~{\sc ii} regions is from -0.1 (optically thin) to 2 (optically thick), though at 7 mm, few sources are optically thick. The largest known spectral index values in sources in Sgr B2 Main are the ``rising spectral index'' sources with $\alpha\sim$1.0 (Gaume et al. 1995, De Pree et al. 1998). But, using these two extreme spectral index values (-0.1 and 2), and the range of observed frequencies, we can determine that in Sgr B2 Main, observations at different frequencies (45.4 GHz in 2003 and 44.2 GHz in 2012) would result in flux density differences at the $<$1\% level for $\alpha\sim$-0.1 (optically thin source) and at the $<$6\% level for $\alpha\sim$2. In Sgr B2 North, observations at different frequencies (45.7 GHz in 2001 and 44.2 GHz in 2012) would result in flux density differences at the $<$1\% level for $\alpha\sim$-0.1 (optically thin source) and at the $<$7\% level for $\alpha\sim$2. 

Thus, we conclude that our observed flux density changes in F3c-Main (flux density increase of $\sim$29\%) and K2-North (flux density decrease of $\sim$27\%) cannot be the result of observations made at slightly different frequencies. The detected decrease in flux density of K2-North, for example, is a factor of $>$4 larger than one would expect from a decrease resulting from a change introduced by observations made at slightly different frequencies.

\section{DISCUSSION}
Galv\'an-Madrid et al.\ (2011) used simulations by Peters et al (2010a, 2010b, 2011) to predict that an individual source observed at 2 cm has a 5\% chance of experiencing
more than a 50\% change in flux density over a 20 year  baseline and a 14\% chance of a change larger than 10\%. 
Consequently, we also expect variations of the observed 7mm flux. However, the actual percentages are highly model dependent, although they are likely of the correct order of magnitude. Furthermore, they will depend on whether the region is optically thick or optically thin, and on the relative contribution of thermal emission from dust.
%In an optically thick region, we might expect to see greater variation at 7 mm than the models predict. In an optically thin region, there should be comparable flux density variations at both wavelengths. Since thermal dust emission is a larger contributor at 7 mm, we expect that in dusty regions, the variations at 7 mm might be less, since the potentially variable 7 mm emission is superimposed on a constant dust emission. 
Since the optical depth of the free-free emission decreases with increasing wavelength, the predictions for the observed variations at 7 mm should be smaller than those estimated 
by Galvan-Madrid et al. (2011) at 2 cm. Furthermore, dust emission contributes more at 7 mm than at 2 cm. These two effects act together to decrease the possible variability at 7mm.
In addition, these two competing factors are likely to be dominated by the sensitivity of the observations. As a result, we expected the ``flickering'' effect might be less at 7 mm than predicted at 2 cm, simply because of the lower signal to noise in the archival data (2003-Main and 2001-North).

{\it Comparison with 1.3 cm Results: }
We find significant changes in flux density of the 7 mm continuum emission from two sources in our sample: K2-North (between 2001 and 2012) and F3c-Main (between 2003 and 2012). In comparison, at 1.3 cm, we detected four sources in the same field that changed in peak intensity over 23 years (between 1989 and 2012; De Pree et al. 2014). K3-North was observed to decrease in peak intensity by 7\%. Sources F1-Main, F3-Main and F1a-Main were detected to increase in peak intensity by 7\%, 5\% and 16\% respectively over the same period.

%F3-Main is remarkable, because the 7mm flux density decreases while the 1.3 cm flux density increases. However, we note that the decrease at 7 mm is measured over a shorter time baseline of 15 yr (from 1997 to 2012) compared to
%23 yr (1989 to 2012) at 1.3 cm. Assuming that the 1.3 cm and 7 mm peak intensities are tightly correlated, the flux density fluctuations reported in this paper and De Pree et al. (2014) are consistent as long as: (1) K2-North increased in flux density from 1989 to 2001 and then decreased until 2012, (2) K3-North decreased in flux density from 1989 to 2001, and remained constant until 2012, (3) F3-Main increased in flux density from 1989 to 1997 and  decreased until 2012, and (4) F1-Main increased in flux density from 1989 to 1997, and remained constant until 2012.

These requirements suggest that flux density variations may occur in a small subset ($<$10\%) of sources on timescales of only a few years. Indeed, such rapid changes have already been found in Orion BN/KL (Gomez et al. 2008; Rivilla et al. 2015) and W3(OH) (Dzib et al. 2013). These recent observations, and our observations of Sgr B2 Main and North indicate that short timescale monitoring of source flux densities in crowded fields of UC and HC H~{\sc ii} regions may identify other variable sources. Our difficulty in isolating and fitting the sources in Main in the current investigation indicates that the technique used at 1.3 cm (precisely matched configurations and bandwidths) makes the job of comparing flux densities between two epochs much simpler.

{\it Submillimeter Array (SMA) 850 \micron~Observations: } Qin et al. (2011) have published SMA observations of the 850 $\micron$ continuum in Sgr B2 Main and Sgr B2 North (Sgr B2-M and Sgr B2 N in their paper). They have reported detections of 2 peaks in Sgr B2 North and 12 in Sgr B2 Main, and discussed the coincidence between their continuum peaks and the known source positions of ultracompact regions in Sgr B2 Main and Sgr B2 North. The brightest continuum peak [SgrB2(M)-SMA1, 2.39$\pm$0.138 Jy beam$^{-1}$] is located near the position of F3c-Main, one of the two sources reported in this paper to change in flux density at 7 mm between 2003 and 2012. Their second brightest continuum source [SgrB2(N)-SMA1, 1.79$\pm$0.039 Jy beam$^{-1}$] lies at the position of K2-North, the other source in which a significant change in 7 mm peak intensity and flux density has been recorded. These observations are consistent with the theoretical suggestion that flickering is caused by strong accretion flows (Peters et al. 2010a).

{\it New Source K7-North: } We have detected a new continuum source to the north of the well known K1-K3 sources at RA 17h 47m 19.895s, Dec -28$^o$ 22\arcmin 13\farcs47. The location of this source has not been previously published to our knowledge, though Qin et al (2011) mention that a nearby source was detected in the 7 mm continuum observations reported by Rolffs et al (2011). The source (which we designate K7-North) is visible in both the 2012 data presented here, and the 2001 data of Mart\'in-Pintado (2015). The source parameters at both epochs are presented in Table 3, and flux density is constant over the 11 year time span observed (2001 to 2012). This source was not reported in the 1.3 cm observations of Gaume et al (1995), most likely because it is located in the midst of diffuse, patchy emission associated with two shell-like sources (K5-North and K6-North). The current high resolution 7 mm image (North-2012) has filtered out most of the diffuse emission from the shells, highlighting this point source. K7-North lies within 0\farcs2 in both RA and Dec from the position of Sgr B2(N)-SMA2, one of two continuum peaks in Sgr B2 North detected in the 850 $\micron$ continuum (Qin et al 2011).

\section{CONCLUSIONS}
New 7 mm continuum observations with the VLA indicate that two of the ultracompact H~{\sc ii}  regions in Sgr B2 (F3c-Main and K2-North) have experienced changes in flux density over short timescales (9 yr and 11 yr respectively). The detection of source variability (and in particular decreases in both peak intensity and flux density in K2-North) support the hypothesis that UC and HC H~{\sc ii} regions may be located in dynamically and morphologically complex and time-variable environments during the first 100,000 yr of their evolution, producing short timescale variability. We are undertaking similar observations of other crowded massive star forming regions to further test this hypothesis.

\acknowledgements{The authors thank the anonymous referee for helpful comments and suggestions that improved the paper. The authors also thank J. Mart\'in-Pintado for allowing us to  utilize his 2001 7 mm observations of Sgr B2 North. CDP and M-MML were supported by NSF grant AST-1211460. M-MML was additionally supported by NSF grant AST-1109395 and the Humboldt Foundation. TP acknowledges financial support through a Forschungskredit of the University of Z\"{u}rich, grant no. FK-13-112, and from the DFG priority program 1537 `Physics of the interstellar medium'. RSK acknowledges support from the Deutsche Forschungsgemeinschaft (DFG) via the SFB 881 ÒThe Milky Way SystemÓ (sub-projects B1, B2, B5 and B8) , and the SPP (priority program) 1573, ÒPhysics of the ISMÓ. He also thanks the European Research Council for significant funding under the European CommunityÕs Seventh Framework Programme (FP7/2007-2013) via the ERC Advanced Grant ÓSTARLIGHTÓ (project number 339177). RG-M acknowledges support from program UNAM-DGAPA-PAPIIT IA101715.}

\newpage
\begin{deluxetable}{lcc}
\tabletypesize{\footnotesize}
\tablenum{1}
\tablecolumns{3}
\tablecaption{Observing Parameters at 7 mm (2012)}
\tablehead{
\bf{Parameter} & \bf{Sgr B2 Main (Main-2012)} & \bf{Sgr B2 North (North-2012)}
}
\startdata
Program & 11B-058 & 11B-058 \\
Observe Date & 21 Sep. 2012 & 28 Oct. 2012\\
Configuration & BnA$\rightarrow$A\tablenotemark{a} & A \\
Observing Frequencies (GHz) & 42.9, 45.4 & 42.9, 45.4 \\
Averaged Frequency (GHz) & 44.2 & 44.2 \\
Total Observing Time (Time on source) & 4h (2h 30m) & 4h (2h 30m) \\
Flux Cal. & J1331+305 & J1331+305 \\
Phase Cal. & J1733-130 & J1733-130 \\
Bandpass Cal. & J1733-130 & J1733-130 \\
Continuum RMS noise (mJy/beam) & 0.14 & 0.16 \\
Synthesized Beam & 0$\farcs12\times0\farcs10, 74^{o}$ & 0$\farcs11\times0\farcs05, 3^{o}$ \\
Total Continuum Bandwidth (MHz) & 1024 & 1024\\
%\\
%\it{Line Observing Parameters}\\
%Bandwidth - RRLs (MHz) & 64 MHz & 64 MHz \\
%Bandwidth - RRLs (km/s) & 448 & 448 \\
%Number of Channels & 128 & 128\\
%Averaged Channel Separation (MHz) & 0.501 & 0.501 \\
%Averaged Channel Separation (km/s) & 3.5 & 3.5 \\
%Channel RMS noise (mJy/beam) & 2.2 & 1.3 \\
%
%\\
\enddata
\tablenotetext{a}{These observations were made while the VLA was being moved from the BnA to the A configuration}
\end{deluxetable}

\newpage

\begin{deluxetable}{lccccc}
\tabletypesize{\footnotesize}
\tablenum{2a}
\tablecolumns{6}
\tablecaption{Sgr B2 Main: 7mm  Radio Continuum Parameters (44.2 GHz)}
\tablehead{
Source & R.A  & Dec.  & Size, PA & Peak Intensity & $S_\nu$ \\
Name & 17 h 47 m & -28$^{\circ}$ 23$\arcmin$            &$\theta_{maj.}\times\theta_{min.}, BPA^{\circ}$     &      [mJy beam$^{-1}$]         &    [mJy] \\
&  ($\pm$0.01 s) & ($\pm0\farcs$1) & &   ($\pm$0.2)  & \\
& [1] & [2] & [3] & [4] & [5]
}
\startdata
B&   19.94 s &  03$\farcs$0 & $\sim0\farcs$9  & 10.9 & 80$\pm$8 \\

B10.06&    19.87 s &   01$\farcs$2 & 0$\farcs29\times0\farcs18, 55^{\circ}$  & 2.0 & 9$\pm$0.9 \\
\\
%D&   20.074 s &   12\farcs76& 0\farcs28$\times$0\farcs10, 29$^{\circ}$  & 0.003 & 0.010\pm$0.002$ & ... & ...\\
%\\
E&   20.11  s &    08$\farcs$8 & $\sim0\farcs$65 & 4.8 & 63$\pm$6 \\
\\
F1a& 20.12 s &   03\farcs6 & 0\farcs18$\times$0\farcs15, 54$^{\circ}$ & 30.2 & 71$\pm$7 
\\ 
F1c& 20.13 s &   03\farcs8 & 0\farcs23$\times$0\farcs19, 165$^{\circ}$  & 25.6 & 92$\pm$9
\\
F1d& 20.13 s &   04\farcs2 & 0\farcs15$\times$0\farcs13, 145$^{\circ}$ & 9.1 & 15$\pm$2
\\
F1f& 20.15 s &   03\farcs9 & 0\farcs32$\times$0\farcs16, 67$^{\circ}$  & 34 & 153$\pm$15 
\\
F1h& 20.12s &   04\farcs2 & 0\farcs14$\times$0\farcs10, 59$^{\circ}$  &10.2 & 16$\pm$1 
\\
\\

F2&  20.18 s &   03\farcs5 & 0\farcs24$\times$0\farcs23, 76$^{\circ}$ & 12.2 & 56$\pm6$ \\
\\

F3a& 20.17 s &   04\farcs9 & 0\farcs22$\times$0\farcs15, 74$^{\circ}$  & 13.7 & 38$\pm$4 \\
F3c&  20.17 s &   04\farcs4 & 0\farcs24$\times$0\farcs16, 112$^{\circ}$ & 33 & 107$\pm$11 \\
F3d&  20.18 s &   04\farcs5 & $\sim$0\farcs35  & 51.4 & 560$\pm$60 \\
F3e& 20.23 s &   04\farcs6 & 0\farcs22$\times$0\farcs19, 156$^{\circ}$  & 5.7 & 21$\pm$2 
\\ \\
F4 & 20.22 s &   04\farcs2 & 0\farcs25$\times$0\farcs17, 139$^{\circ}$ & 25.4 & 92$\pm$9 \\
\\
F10.27& 20.09 s&  05\farcs2 & 0\farcs12$\times$0\farcs10, 76$^{\circ}$ & 3.8 & 4$\pm$0.4 \\
F10.37& 20.19 s&  05\farcs8 & 0\farcs15$\times$0\farcs12, 101$^{\circ}$ & 23.6 & 38$\pm$4 \\
F10.39& 20.21 s &  06\farcs5 & 0\farcs13$\times$0\farcs11, 83$^{\circ}$ & 15.1 & 18$\pm$2 \\

\\
G&   20.30 s &  02\farcs9 & 0\farcs26$\times$0\farcs19, 54$^{\circ}$  & 25.5 & 109$\pm$11 \\
G10.44&   20.26 s &  03\farcs2 & 0\farcs12$\times$0\farcs10, 64$^{\circ}$  & 6.0 & 6$\pm$1\\

\\
I&   20.39 s &   05\farcs0& $\sim$2\farcs9  & 4.1 & 318$\pm$32 
\\
I10.52&   20.34 s &   08\farcs00 & 0\farcs16$\times$0\farcs14, 51$^{\circ}$  & 4.1 & 7$\pm$0.7 \\  
\enddata
\end{deluxetable}

\begin{deluxetable}{lccccc}
\tabletypesize{\footnotesize}

\tablenum{2b}
\tablecolumns{6}
\tablecaption{Sgr B2 North: 7mm  Radio Continuum Parameters (44.2 GHz)}
 
\tablehead{
Source & R.A  & Dec.  & Size, PA & Peak Intensity & $S_\nu$ \\
Name & 17 h 47 m & -28$^{\circ}$ 22$\arcmin$            & $\theta_{maj.}\times\theta_{min.}, BPA^{\circ}$     &      [mJy beam$^{-1}$]         &    [mJy] \\
&  ($\pm$0.01 s) & ($\pm$0\farcs1) & &   ($\pm$0.2)  & \\
& [1] & [2] & [3] & [4] & [5]
}

\startdata
K1& 19.77 s& 20\farcs6 & 0$\farcs45\times0\farcs25, 5^{\circ}$ & 7.1 & 135$\pm$14\\

K2& 19.87 s& 18\farcs5 & 0$\farcs22\times0\farcs10, 163^{\circ}$ & 22.3 & 80$\pm$8\\

K3& 19.21 s& 17\farcs2 & 0$\farcs29\times0\farcs27, 141^{\circ}$ & 18.7 & 242$\pm$24 \\

K4 & 20.02 s& 4\farcs8 & $\sim$0\farcs9\tablenotemark{a}                     & 2.4   & 124$\pm$12 \\

K7& 19.90 s&  13\farcs5 & 0$\farcs14\times0\farcs11, 147^{\circ}$    & 11.9   &30$\pm$3\\
\enddata
\tablenotetext{a}{Source size measurement for the shell-like source K4-North is described in the text.}
\end{deluxetable}

\newpage

\begin{deluxetable}{lccccccc}
\tabletypesize{\tiny}

\tablenum{3}
\tablecolumns{9}
\tablecaption{Sgr B2 North 7mm  Radio Continuum Parameters - matched \it{uv-}-coverage}
\scriptsize
\tablehead{
Source & R.A  & Dec.  & Size, PA & Peak Intensity & $S_\nu$ & Peak Intensity & Flux Density\\
Name & 17 h 47 m & -28$^{\circ}$ 22$\arcmin$            & $\theta_{maj.}\times\theta_{min.}, BPA^{\circ}$     &      [mJy beam$^{-1}$]         &    [mJy]   & Ratio & Ratio\\
($\theta_{beam}$) & ($\pm$0.01 s) & ($\pm$0\farcs1) & &  & & (2012/2001) & (2012/2001) \\
& [1] & [2] & [3] & [4] & [5] & [6] & [7]
}    
  
\startdata

\bf{North-2012m} \\
($0\farcs13\times0\farcs10, BPA=0^o$) & & & & ($\pm0.3 mJy beam^{-1}$) 
\\
K1& 19.77 s& 20\farcs6 & 0$\farcs47\times0\farcs30, 40^{\circ}$& 13.6 & 143$\pm$14 & 0.99$\pm$0.15 & 0.95$\pm$0.13\\

K2& 19.87 s& 18\farcs4 & 0$\farcs30\times0\farcs15, 157^{\circ}$& 28.9 & 103$\pm$10 & 0.71$\pm$0.05 & 0.73$\pm$0.10\\

K3& 19.90 s& 17\farcs2 & 0$\farcs30\times0\farcs28, 139^{\circ}$& 37.5 & 239$\pm$24& 1.06$\pm$0.06 & 1.11$\pm$0.15\\

K4 & 19.99 s& 04\farcs8 & $\sim0\farcs9$\tablenotemark{a}                     & 4.1   & 139$\pm$14 &  ... & ..\\

K7& 19.90 s&  13\farcs5 & 0$\farcs16\times0\farcs14, 130^{\circ}$    & 18.0    &31$\pm$3 & 0.78$\pm$0.09 & 1.03$\pm$0.14\\

\\
\bf{North-2001}\\
($0\farcs13\times0\farcs10, BPA=0^o$) & & & & ($\pm$2.0 mJy beam$^{-1}$)
\\
K1&  19.79 s&  20\farcs5 & 0\farcs51$\times0\farcs28, 15^{\circ}$ & 13.8 & 151$\pm$15&  ... & ... \\

K2& 19.89 s& 18\farcs3 & 0\farcs30$\times0\farcs15,  163^{\circ}$ & 40.7 & 142$\pm$14& ... & ...\\

K3& 19.91 s&  17\farcs0& 0\farcs30$\times0\farcs27, 106^{\circ}$ & 35.1 & 216$\pm$22& ... & ...\\

K4 & ...  &  ... & ...                            & ... & ... & ... & ...\\

K7& 19.91 s&  13\farcs3 & 0\farcs14$\times0\farcs12, 165^{\circ}$    & 23.2  &30$\pm$3   & ... & ...  \\
\enddata

%\tablenotetext{a}{Source size measurements were made with the AIPS task JMFIT. For source K4 (that could not be well fitted with a 2D gaussian) the AIPS task IMSTAT was used to measure peak flux density and integrated flux density within a region.}
\tablenotetext{a}{Source size measurement for the shell-like source K4-North is described in the text.}
%\tablenotetext{b}{The lower native resolution of the 2001 observations make these data more sensitive to the diffuse emission in this more extended source, resulting in an apparent decrease in peak brightness, albeit with low significance relative to the decrease in the ultracompact source K2.}

\end{deluxetable}
%\end{landscape}

\newpage

\begin{figure}[t]
 \figurenum{1}
\begin{center}
  \includegraphics[width=1.0\textwidth]{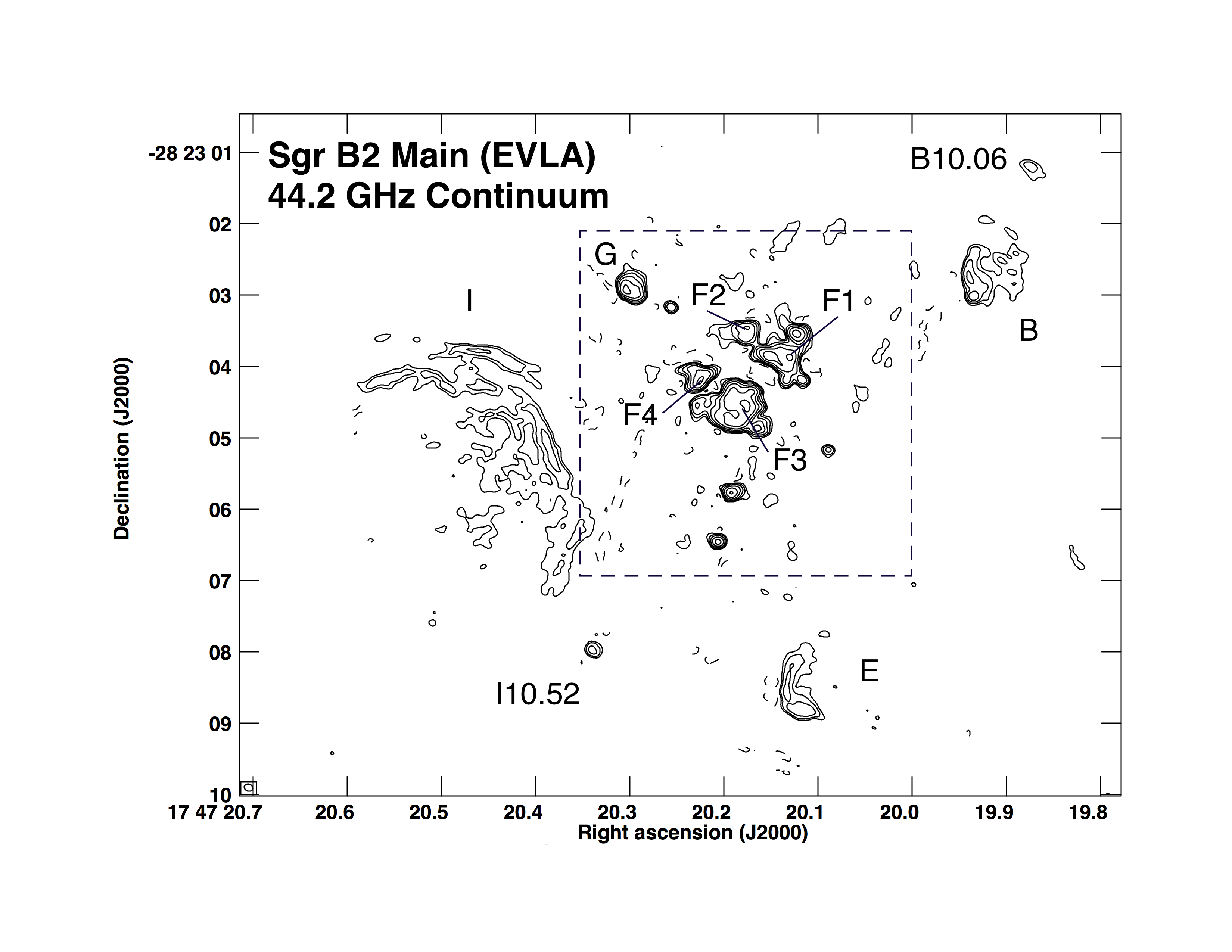}
  \end{center}
\caption{The central region of Sgr B2 Main as observed at 44.2 GHz with the EVLA in 2012 in the BnA configuration. Restoring beam (indicated in the lower left hand corner of the image) is 0\farcs12$\times$0\farcs10, BPA=74$^o$. First positive and negative contours are at the 5$\sigma$ level in the 2012 continuum (0.7 mJy beam$^{-1}$). Successive positive contours are at 2, 4, 8, 16 and 32 times the 5$\sigma$ level. Sources are labelled using the naming conventions in Gaume et al. (1995) and De Pree et al. (1998). The dashed box surrounding the F and G sources indicates the region of detail in Fig. 4. Names of sub sources located in this box are given in Fig. 4.}

%(contours, $\theta$_{beam} = 0\farcs12$\times$0\farcs10, BPA=74$^o$) and convolved resolution (grayscale, $\theta$_{beam} = 0\farcs14$\times$0\farcs12, BPA=70$^o$). Sources are labelled as indicated in Gaume et al. (1995) and De Pree et al (2011). First positive and negative contours are at the 5$\sigma$ level in the 2012 continuum (1.25 mJy beam$^{-1}$), successive positive contours are at 2, 4, 8, 16, and 32 times the 5$\sigma$ contour. (b) Sgr B2 Main at 45.4 GHz (Main-2003) at convolved resolution (contours, $\theta$_{beam} = 0\farcs14$\times$0\farcs12, BPA=70$^o$). First positive and negative contours are at the 3$\sigma$ level in the 2003 continuum (3.9 mJy beam$^{-1}$), successive positive contours are at 2, 4, 8, 16, and 32 times the 3$\sigma$ contour. (c) Sgr B2 Main at 43.3 GHz (Main-1997) at convolved resolution (contours, $\theta$_{beam} = 0\farcs14$\times$0\farcs12, BPA=70$^o$). First positive and negative contours are at the 3$\sigma$ level in the 1997 continuum (6.00 mJy beam$^{-1}$), successive positive contours are at 2, 4, 8, 16, and 32 times the 3$\sigma$ contour. The beam size for each image is indicated in the lower left hand corner of each image.}}

\end{figure}

\begin{figure}[t]
 \figurenum{2}
\begin{center}
  \includegraphics[width=0.8\textwidth]{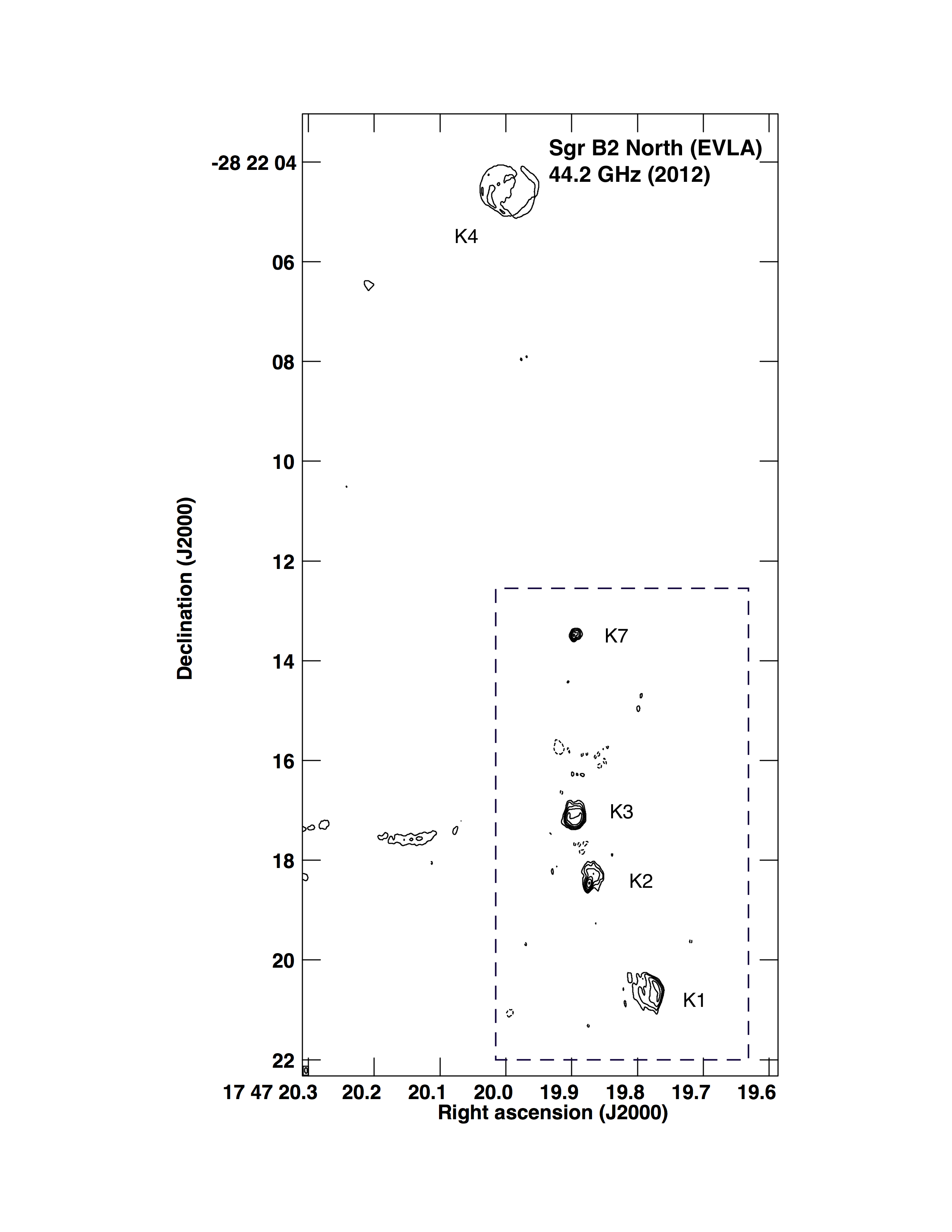}
  \end{center}
  \caption{Sgr B2 North as observed at 44.2 GHz with the EVLA in 2012 in the A configuration. Restoring beam (indicated in the lower left hand corner of the image) is 0\farcs11$\times$0\farcs05, BPA=3$^o$. First positive and negative contours are at the 5$\sigma$ level in the continuum (0.8 mJy beam$^{-1}$). Successive positive contours are at 2, 4, 8, 16 and 32 times the 5$\sigma$ level. Sources are labelled using the naming conventions in Gaume et al. (1995). The dashed box surrounding the K1 to K7 sources indicates the region of detail in Fig. 3. K7 is a newly detected source at 7 mm.}
  
%\caption{\footnotesize{7 mm (44.2 GHz) continuum image of Sgr B2 Main (2012), observed during the BnA$\rightarrow$A move of the VLA ($\theta$_{beam} = 0\farcs12$\times$0\farcs10, BPA=74$^o$). First contour is at 5$\sigma$ (1.25 mJy beam$^{-1}$). Successive positive contours are at 1.4, 2, 2.8, 4, 5.6, 8, 11.2, 16, and 32 times the first contour level. Sources are labelled with the naming convention of Gaume et al. (1995). Inset (lower right) shows the 1997 VLA 7 mm (43.3 GHz) continuum and the 2012-1997 difference map (contours) in the region highlighted in Fig. 1. In the difference map, first contours are at the 10$\sigma$ level of the 1997 image. Successive negative contours are at the 15 and 20$\sigma$ levels. Sub sources of F1-F4 are indicated in Fig. 1. The beam size is indicated in the lower left hand corner of each image.}}
\end{figure}

\begin{figure}[t]
 \figurenum{3}
\begin{center}
\includegraphics[width=0.85\textwidth]{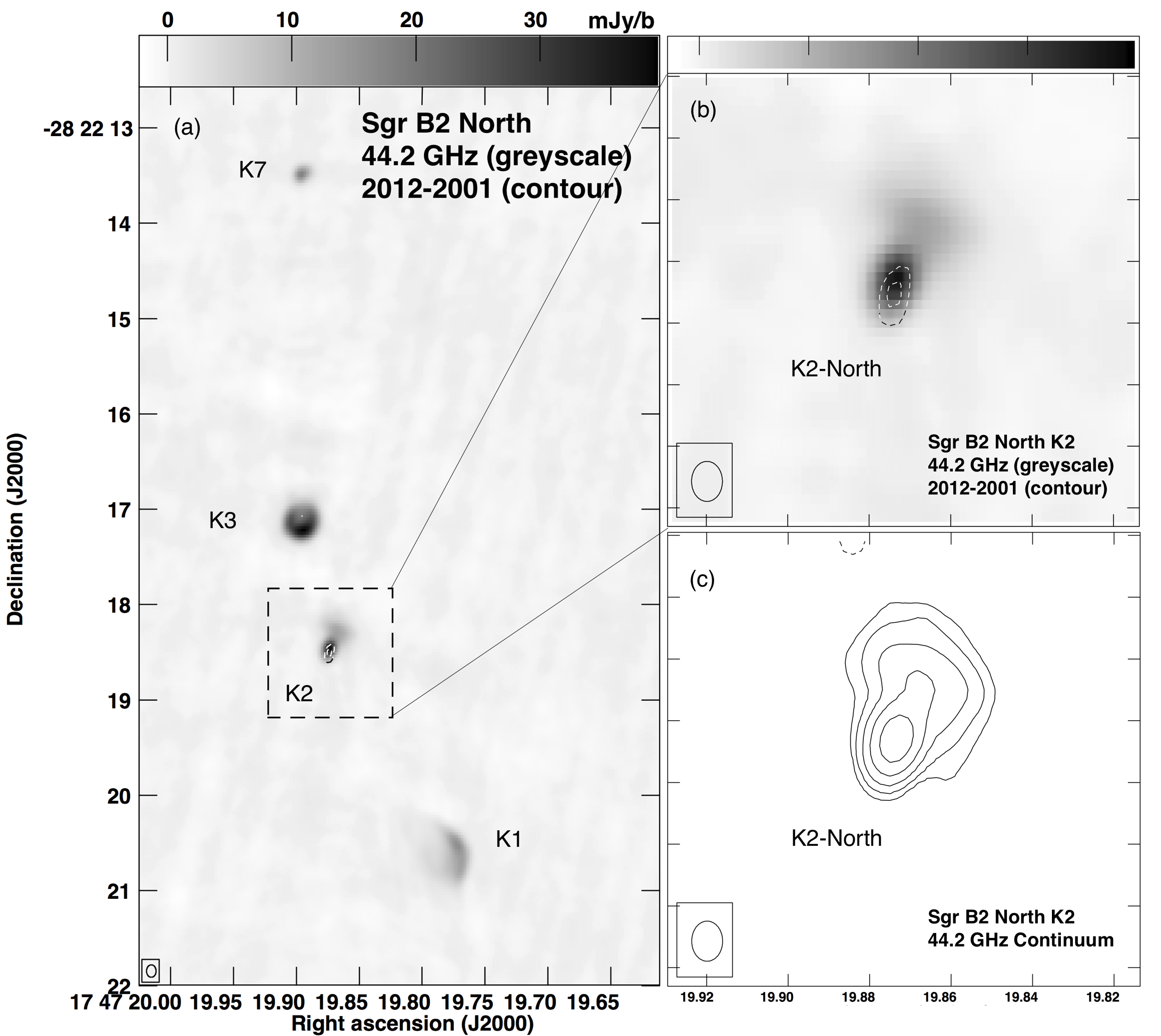}
\end{center}
\caption{\footnotesize{Flux density changes in Sgr B2 North (a) Greyscale shows the continuum emission in the region of detail indicated in Fig. 2 from Sgr B2 North in 2012 with the difference image between 2012 and 2001 overlaid (contours). The uv-coverage, bandwidth and restoring beam were matched between these two epochs as described in the text. Both the images used to make the difference overlay were made with a restoring beam of $\theta_{beam}$ = 0\farcs13$\times$0\farcs10, BPA = 0$^o$. In the difference image (contours), the negative contours are at the 5$\sigma$ (10 mJy beam$^{-1}$) and 7.5$\sigma$ (15 mJy beam$^{-1}$) levels of the 2001 image . The beam size is indicated in the lower left hand corner of the image. (b) shows a larger view of the region of detail indicated in (a), showing the decrease in flux density detected in K2-North. (c) shows the continuum emission from K2-North in the matched-resolution data (North-2012m). First positive and negative contours are at 5$\sigma$ in the 2012 image (1.4 mJy beam$^{-1}$).}}
\end{figure}

\begin{figure}[t]
\figurenum{4}
\begin{center}
\includegraphics[width=1.0\textwidth]{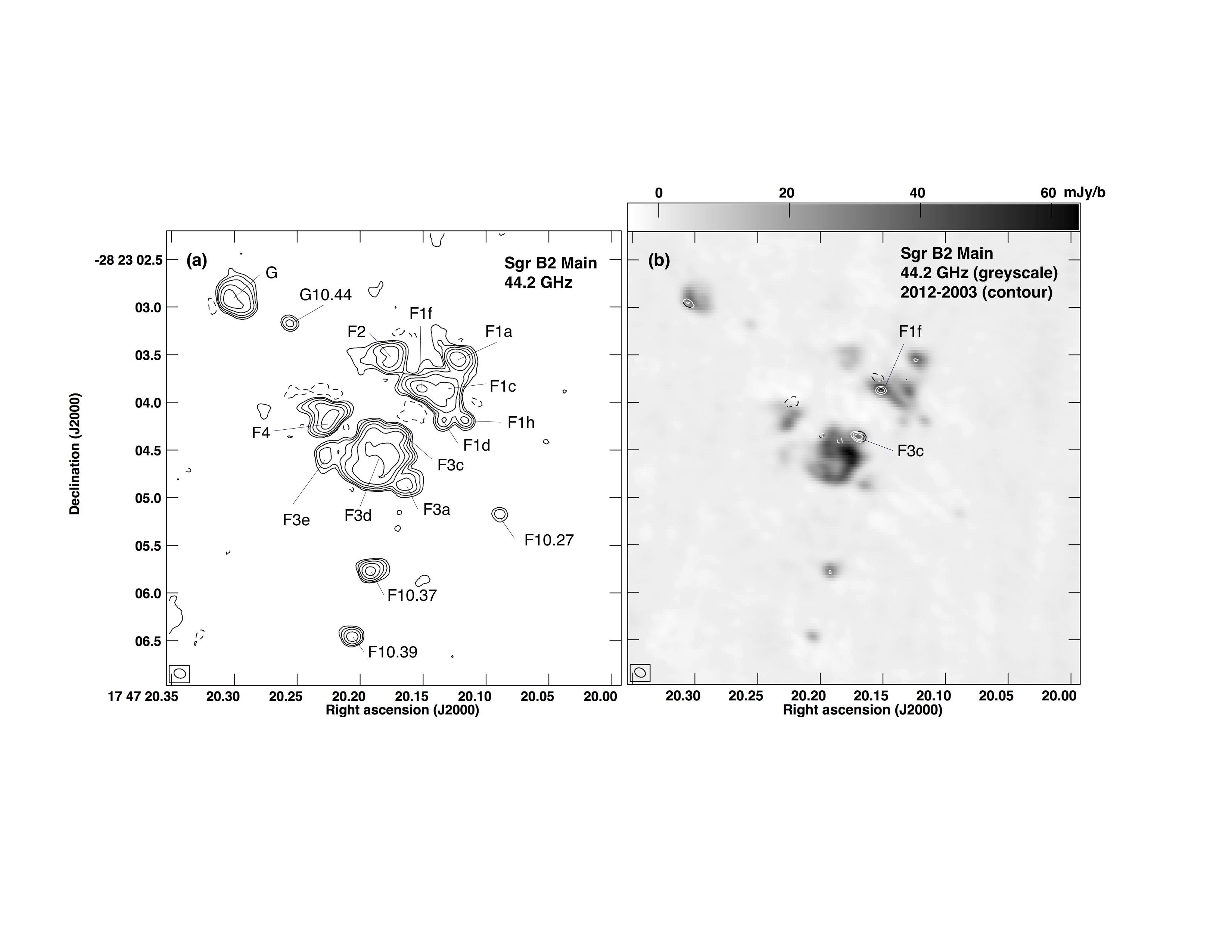}
\end{center}
\caption{\footnotesize{Both (a) and (b) show the same region of detail from the image of Sgr B2 Main in Fig. 1. (a) Contours show continuum emission from Sgr B2 Main in 2012. First negative and positive contours are at the 7$\sigma$ (1 mJy beam$^{-1}$) level in the 2012 image. Successive positive contours are at 2, 4, 8, 16 and 32 times the 7$\sigma$ level. A slightly higher contour cutoff was chosen for this image to make source labels easier to read. Sources are labeled as named in De Pree et al. (1998). (b) Image shows 2012-Main (greyscale) with the difference image between 2012 and 2003 overlaid (contours). The uv-coverage, bandwidth and restoring beam were matched between these two epochs as described in the text. Both the images used to make the difference overlay were made with a restoring beam of $\theta_{beam}$ = 0\farcs12$\times$0\farcs09, BPA = 60$^o$. In the difference image (contours), the first negative and positive contours are at the 5$\sigma$ level of the 2003 image (4 mJy beam$^{-1}$). The beam size is indicated in the lower left hand corner of each image.}}
\end{figure}

%\begin{figure}[t]
% \figurenum{3b}
%\begin{center}
 % \includegraphics[width=0.85\textwidth]{Fig3b.jpg}
%  \end{center}
%\caption{Comparison between the 7 mm flux density in 2001 and 2012 for the detected continuum sources in Sgr B2 North. Note that most sources are constant in integrated flux density, while source K2 experiences a significant decrease.}
%\end{figure}

\begin{figure}[t]
\figurenum{5}
\begin{center}
\includegraphics[width=0.95\textwidth]{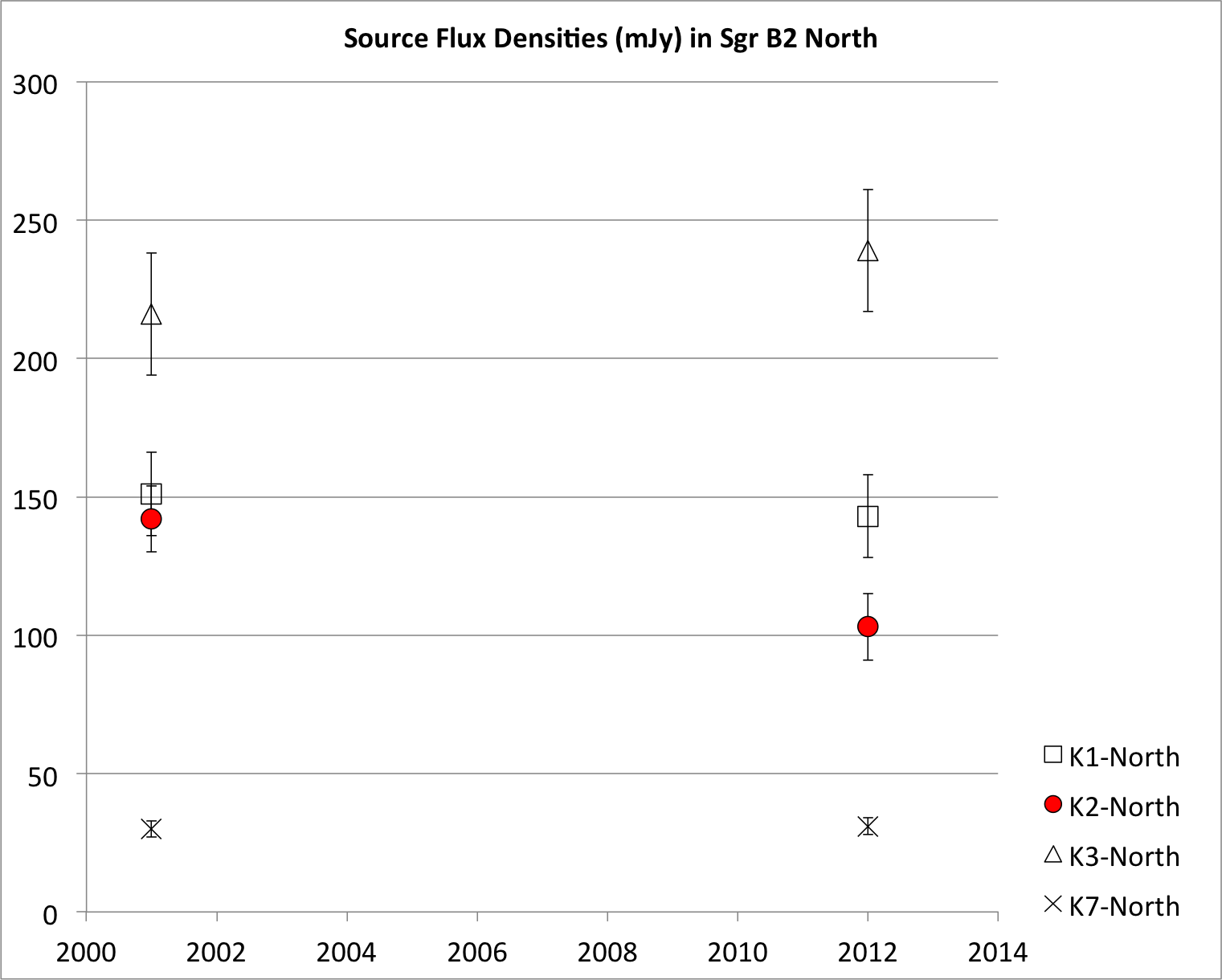}
\end{center}
\caption{\footnotesize{Source flux densities for the four (4) continuum sources detected in both 2001 and 2012 toward Sgr B2 North. Fitted flux density values for K1-North (open square), K2-North (red circle), K3-North (open triangle) and K7-North (x) are shown for 2001 and 2012. Error bars are at 10\% level for each flux density value.}}
\end{figure}

%\begin{figure}[t]
 %\figurenum{4}
%\begin{center}
 % \includegraphics[width=0.9\textwidth]{MainRRL2.png}
 % \end{center}
%\caption{Gaussian fits to the averaged H52$\alpha$ and H53$\alpha$ (7 mm) Radio Recombination Line in Sgr B2 Main. Sources with line-to-continuum ratio detections above the 3$\sigma$ level are shown. For each source, integrated line to continuum ratio (red crosses), Gaussian fit (blue line) and residual (dotted green line) are shown.}
%\end{figure}

%\begin{figure}[t]
 %\figurenum{4}
%\begin{center}
  %\includegraphics[width=0.9\textwidth]{MainRRL3.png}
  %\end{center}
%\caption{Gaussian fits to the averaged H52$\alpha$ and H53$\alpha$ (7 mm) Radio Recombination Line in Sgr B2 Main. Sources with line-to-continuum ratio detections above the 3$\sigma$ level are shown. For each source, integrated line to continuum ratio (red crosses), Gaussian fit (blue line) and residual (dotted green line) are shown.}

%\end{figure}

%\begin{figure}[t]
 %\figurenum{5}
%\begin{center}
  %\includegraphics[width=0.9\textwidth]{NorthRRLs.png}
  %\end{center}
%\caption{Gaussian fits to the 7 mm averaged H52/53$\alpha$ Radio Recombination Line in Sgr B2 North. Sources with detections above the 3$\sigma$ level are shown. For each source, integrated line to continuum ratio (red crosses), Gaussian fit (blue line) and residual (dotted green line) are shown.}
%\end{figure}

\end{document}